\documentclass[journal]{IEEEtran}

\def\old#1{}

\usepackage{acronym}
\usepackage{amstext}
\usepackage{url}
\usepackage{graphicx}
\usepackage{epstopdf}
\usepackage{amsmath}
\usepackage{float}
\usepackage{enumerate}
\usepackage{amsfonts}
\usepackage{algorithmic}
\usepackage{algorithm}
\usepackage{cite}
\usepackage{color}

\usepackage{boxedminipage}
\usepackage{listings}
\usepackage{subfigure}
\usepackage{caption2}

\newtheorem{remark}{Remark}
\newtheorem{example}{Example}

\newcommand{\be}{\begin{equation}}
\newcommand{\ee}{\end{equation}}

\providecommand{\boldsymbol}[1]{\mbox{\boldmath $#1$}}

\makeatother

\newcommand{\vc}[1]{\boldsymbol{#1}}
\def\real    { \mathbb{R} }

\acrodef{i.i.d.}{independent and identically distributed}
\acrodef{WLS}{Weighted Least Squares}

\acrodef{CoM}{Concentration of Measure}
\acrodef{i.i.d.}{independent and identically distributed}
\acrodef{LTI}{Linear Time-Invariant}
\acrodef{LTV}{Linear Time-Variant}
\acrodef{LPV}{Linear Parameter-Varying}
\acrodef{RIP}{Restricted Isometry Property}
\acrodef{SVD}{Singular Value Decomposition}
\acrodef{CS}{Compressive Sensing}
\acrodef{DSP}{Digital Signal Processing}
\acrodef{CSI}{Compressive System Identification}
\acrodef{CTI}{Compressive Topology Identification}
\acrodef{CBD}{Compressive Binary Detection}
\acrodef{OMP}{Orthogonal Matching Pursuit}
\acrodef{MP}{Matching Pursuit}
\acrodef{ERC}{Exact Recovery Condition}
\acrodef{BOMP}{Block Orthogonal Matching Pursuit}
\acrodef{COMP}{Clustered Orthogonal Matching Pursuit}
\acrodef{CoSaMP}{Compressive Sampling Matching Pursuit}
\acrodef{KKT}{Karush-Kuhn-Tucker}

\acrodef{FIR}{Finite Impulse Response}
\acrodef{DFT}{Discrete Fourier Transform}
\acrodef{DCT}{Discrete Cosine Transform}
\acrodef{JL}{Johnson-Lindenstrauss}
\acrodef{ROC}{Receiver Operating Curve}
\acrodef{NP}{Neyman-Pearson}
\acrodef{ARX}{Auto Regressive with eXternal input} 
\acrodef{MISO}{Multi-Input Single-Output}
\acrodef{SISO}{Single-Input Single-Output}
\acrodef{MIMO}{Multi-Input Multi-Output}
\acrodef{BP}{Basis Pursuit}
\acrodef{LASSO}{Least Absolute Shrinkage and Selection Operator}
\acrodef{GLASSO}{Group LASSO}
\acrodef{NNG}{Non-Negative Garrote}
\acrodef{LARS}{Least Angle Regression}
\acrodef{I/O}{Input/Output}
\acrodef{ME}{Matrix Estimation}
\acrodef{OD}{Origin-Destination}
\acrodef{CODE}{Compressive Origin-Destination  Estimation}
\acrodef{AVI}{Automatic Vehicle Identification}
\acrodef{ETC}{Electronic Toll Collection}
\acrodef{VMT}{Vehicle-Miles Traveled}
\acrodef{CDF}{Cumulative Distribution Function}

\begin{document}
%\begin{titlepage}
\title{Compressive Origin-Destination  Estimation}
%Borhan M. Sanandaj$\text{i}^{*}$ 
\author{Borhan M. Sanandaji and Pravin Varaiya
%\textsuperscript{$*$}Corresponding author.
\thanks{B. M. Sanandaji and P. Varaiya are with the Department of Electrical Engineering and Computer Sciences, University of California, Berkeley, CA 94720 USA 
(e-mail: \{sanandaji,varaiya\}@berkeley.edu).
}
%
%\thanks{\textsuperscript{$\pi$}
\thanks{
This research was funded in part by the California Department of Transportation under the Connected Corridors program. We are grateful to Alex A. Kurzhanskiy for  insights
and to Keir Opie and Vassili Alexiadis for the data from East Providence.
}
}

\date{\today}
\maketitle

\begin{abstract}

The paper presents an approach to estimate Origin-Destination (OD) flows and their path splits, based on traffic counts on links in the network. The approach called Compressive Origin-Destination  Estimation (CODE) is inspired by Compressive Sensing (CS) techniques. Even though the estimation problem is underdetermined, CODE recovers the unknown variables exactly when the number of  alternative paths for each OD pair is small. Noiseless, noisy, and weighted versions of CODE are illustrated for synthetic networks, and with real data for a small region in East Providence. CODE's versatility is suggested by its use to estimate the number of vehicles and the Vehicle-Miles Traveled (VMT) using link counts.

\end{abstract}
%\thispagestyle{empty}
%\end{titlepage}

\section{Introduction}
%%%%%%%%%%%%%%%%%%%%%%%%%%%%%%%%%%%%%%%%%%%%%%%%%%%%%%%%%%%%%
A common task in transportation planning is to estimate path allocations, that is the \ac{OD} flows and the path splits for each flow, from traffic counts on individual links in a network\cite{van1980most,cascetta1984estimation,bell1991estimation,yang1992estimation,abrahamsson1998estimation,
hazelton2003some,bert2009thesis,bera2011estimation}. 
(Estimation using tag data from \ac{AVI} and \ac{ETC} is discussed in \cite{van1997dynamic, asakura2000origin,kwon2005real,zhou2006dynamicITS}.)
One may use static or dynamic traffic models~\cite{bierlaire2002total,barcelo2005dynamic,bert2009thesis}. In a static model the flows and link counts are time-independent. In a dynamic model the flows and link counts are time-dependent\cite{van1980most,okutani1984dynamic,cremer1987new,bell1991estimation,zhou2006dynamic,verbas2011time}.

The unknown path allocation vector $\vc{x}$ contains all \ac{OD} pair flows and path splits for each \ac{OD} pair.
The problem is to recover $\vc x \in \real^{N}$ from the link count measurement vector $\vc{y} \in  \real^M$ when $N$ is much larger than $M$. The two vectors are related by
$\vc y = A \vc x$  in which $A$  is the known binary incidence matrix that specifies the links along each path. We approach the  problem supposing that $\vc x$  is sparse,  i.e., the number $S$ of non-zero entries in $\vc x$ is much smaller than $N$. 
The approach called \ac{CODE} is inspired by \ac{CS}\cite{donoho2006compressed,cands2006compressive,candes2008people} techniques  developed in signal processing for sparse signal recovery. We show that \ac{CODE} recovers the path allocation vector $\vc x$ when it is suitably sparse. 
The main technical novelty of our approach is to formulate the estimation problem as $\ell_1$-recovery of a \textit{sparse} signal 
$x$.

Our motivation behind imposing the sparsity condition on $\vc{x}$ is due to the fact that in a typical urban area there are many possible paths between any given \ac{OD} pair while only a small fraction of these paths are plausible to the travelers based on travel length, travel time, number of turns, etc. \ac{CODE} assumes that the plausible paths between any given \ac{OD} pair is sparse. We give a brief example as to why sparsity is likely. Consider a grid square network with nodes indexed $(m,n)$ from bottom-left to the top-right of the grid. Links only go west to east or south to north. The most bottom-left node (origin) is (0,0) and the most top-right node (destination) is $(N/2,N/2)$. There are $N \choose N/2$ possible paths that connects origin (0,0) to destination $(N/2,N/2)$. For example, for $N=50$ there exist $1.2641 \times 10^{+14}$ possible paths. While each path traverses N links (i.e., all paths have the same length), it may contain different number of turns. For example, there only exist two paths with just 1 turn. The fraction of paths that make at most $\alpha N$ turns is bounded using the Hoeffding's inequality by $\beta = \exp(-2(0.5-\alpha)^2N)$.
Suppose we believe that drivers will not take a route with more  than $0.1N$ turns ($\alpha = 0.1$). For $N = 50$ the fraction of plausible routes with at most $5$ turns is less than $10^{-7}$ of all possible routes. If they won't take routes with more than $0.2N$ turns ($\alpha  = 0.2$), the fraction of all plausible routs with at most $10$ turns is at most $10^{-4}$. The ratio $\beta$ is small and will decrease as N increases, motivating the sparsity assumption on $\vc{x}$.

In section~\ref{sec:setup}, we formulate the  problem for both  static and dynamic  models and introduce the sparse path allocation framework. We list   basic tools of \ac{CS} and $\ell_1$-minimization in section~\ref{sec:cs}. We introduce \ac{CODE} in its noiseless and noisy settings with a small   example in section~\ref{sec:synthetic}. A weighted version of \ac{CODE} is examined in section~\ref{sec:weighted}. In Section~\ref{sec:vmt} we consider the Nguyen-Dupuis network and show how the proposed framework can be used to estimate \ac{VMT}. A  study based on data taken from a traffic area in East Providence is presented in section~\ref{sec:case}. 

%%%%%%%%%%%%%%%%%%%%%%%%%%%%%%%%%%
%%%%%%%%%%%%%%%%%%%%%%%%%%%%%%%%%%
%%%%%%%%%%%%%%%%%%%%%%%%%%%%%%%%%%
%%%%%%%%%%%%%%%%%%%%%%%%%%%%%%%%%%

\section{Problem Setup}
\label{sec:setup}
\begin{figure}[tb]
\centering
\includegraphics[width=0.65\linewidth]{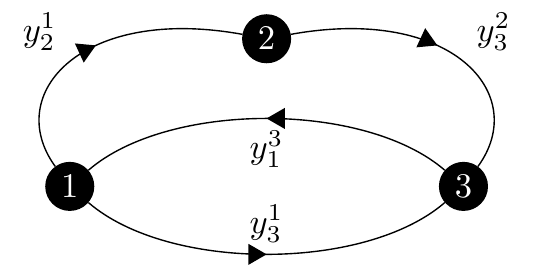}
\caption{Traffic network  with 3 \ac{OD} zones, 4 links, and 7 possible paths associated with 6 possible \ac{OD} pairs. In this example, for each \ac{OD} pair there is only one path except for \ac{OD} pair 1-3 which has 2 alternative paths: $\ell_2^1\ell_3^2$ and $\ell_3^1$.}
\label{fig:network_model}
\end{figure}

Consider a traffic network like in Fig.~\ref{fig:network_model} with nodes or zones indexed  $i, j$ and unidirectional links  $\ell_j^i$  from $i$ to $j$.
An \ac{OD} pair might be  connected by several alternative paths.  For example, from origin $2$ to destination $1$ there is a single path that traverses nodes $2 \rightarrow 3 \rightarrow 1$, while from origin $1$ to destination $3$ there are two paths: $1\rightarrow 3$ 
 and $1 \rightarrow 2 \rightarrow 3$.
The  number of vehicles $y_j^i$ that pass through  link $\ell_j^i$  is measured so, for example,
\[y_2^1 = x^{\ell^1_2} + x^{\ell^1_2 \ell^2_3} + x^{\ell^3_1\ell^1_2} := x_1+x_2+x_7,
\] 
in which  $x^{p_i}$ (also written $x_i$) is the number of vehicles  that take path $p_i$. In this example there are seven possible paths associated with six different \ac{OD} pairs. The vector $y$
of traffic counts is a linear function of the vector $x$ of path flows:

\begin{equation}
% \small
\underbrace{
\begin{bmatrix}
y_2^1\\
y_3^1\\
y_3^2\\
y_1^3
\end{bmatrix}
}_{\vc{y}}
=
\underbrace{
\begin{bmatrix}
1 & 1 & 0 & 0 & 0 & 0 & 1 \\
0 & 0 & 1 & 0 & 0 & 0 & 0 \\
0 & 1 & 0 & 1 & 1 & 0 & 0 \\
0 & 0 & 0 & 1 & 0 & 1 & 1 \\
\end{bmatrix}
}_{A}
\underbrace{
\begin{bmatrix}
x_1\\
x_2\\
x_3\\
x_4\\
x_5\\
x_6\\
x_7\\
\end{bmatrix}
}_{\vc{x}}.
\label{eq:small_network_static}
\end{equation}
%%%%%%%%%%
The matrix element $A_{ij} = 1$ or $0$ accordingly as link $i$ belongs to path $j$ or not.
The goal  is to recover the path allocations $\vc{x}$ from observed link counts $\vc{y}$. Since this is an underdetermined set of of linear equations  unique recovery of the true solution is generally not possible. However we show that under sparsity conditions on the true $\vc{x}$ we can recover a unique solution.  We describe the traffic model next.

\subsection{Static  Model}
\label{sec:staticModel}
There are $K$ \ac{OD} pairs ($k = 1,2,\dots,K$) and $N$ paths ($n=1,2,\dots,N$).  $y^{i}_{j}$ is the vehicle count on link $\ell^i_j$ and $f_k$ is the flow over the $k$th \ac{OD} pair. The  measurement model is
\begin{equation}
y^{i}_{j} = \sum_k \left(\sum_n w_{k,n}a^{i \to j}_{k,n}\right) f_k,
\label{eq:conservation_law_1}
\end{equation}
where $w_{k,n}$ is the fraction of the flow $f_k$ over the $k$th OD pair that takes the $n$th path, and $a^{i \to j}_{k,n} = 1$ if $\ell^i_j$ belongs to the $n$th path in  the $k$th \ac{OD} pair and $= 0$ otherwise~\cite{castillo2008observability}.  The weights or path splits $w_{k,n}$ satisfy 
\begin{equation}
		%\textrm{(weight constraints):} \left\{ \begin{array}{cc} 
			\sum_n w_{k,n} = 1, \;  \;
0 \leq w_{k,n} \leq 1.
 			%\end{array} \right.
		 \label{eq:constraints_on_weights}
\end{equation}
%
% for $k = 1,\dots, K$ and $n = 1,\dots, N$.
%
We reformulate the measurement equation (\ref{eq:conservation_law_1}) as
\begin{equation}
y^{i}_{j} = [a^{i \to j}_{1,1} \dots a^{i \to j}_{k,n} \dots a^{i \to j}_{K,N}]
\underbrace{
\begin{bmatrix}
w_{1,1}f_1 \\ \vdots \\ w_{k,n}f_k \\ \vdots \\ w_{K,N}f_K
\end{bmatrix}
}_{\vc{x} \in \real^N}.
\label{eq:conservation_law_matrix_form_1}
\end{equation}
In (\ref{eq:conservation_law_matrix_form_1}) the dimension of the unknown vector $\vc{x}$ is the number of paths in a network. 

For  Fig.~\ref{fig:network_model}, define $p_1=\ell^1_2, p_2=\ell^1_2 \ell^2_3, p_3=\ell^1_3, p_4=\ell^2_3\ell^3_1, p_5=\ell^2_3, p_6=\ell^3_1, p_7=\ell^3_1 \ell^1_2$. From (\ref{eq:constraints_on_weights}), $w_{1,1} = w_{3,4} = w_{4,5} = w_{5,6} = w_{6,7} =1$ and $w_{2,2}+w_{2,3} = 1$, so the measurement equation becomes
\begin{equation}
		\left\{ \begin{array}{ll} 
	%	& \hspace{-0.25in} \textrm{static link traffic counts:} \\ 
%		& \\
			y^1_2 &= \underbrace{w_{1,1}f_1}_{x_1} + \underbrace{w_{2,2}f_2}_{x_2} + \underbrace{w_{6,7}f_6}_{x_7} =  f_1 + w_{2,2}f_2 + f_6,\\
y^1_3 &= \underbrace{w_{2,3}f_2}_{x_3}=w_{2,3}f_2,\\
y^2_3 &= \underbrace{w_{2,2}f_2}_{x_2}+\underbrace{w_{3,4}f_3}_{x_4}+\underbrace{w_{4,5}f_4}_{x_5}=w_{2,2}f_2+f_3+f_4,\\
y^3_1 &= \underbrace{w_{3,4}f_3}_{x_4}+\underbrace{w_{5,6}f_5}_{x_6}+\underbrace{w_{6,7}f_6}_{x_7}=f_3+f_5+f_6.
			\end{array} \right.
\label{eq:small_network_static_linkflows}
\end{equation}

\begin{remark}
If CODE recovers $\vc{x}$, one  also obtains the \ac{OD} flows $f_k$ and path splits $w_{k,n}$.
For example, given $x_2$ and $x_3$, we can recover $f_2$ because $x_2 + x_3 = w_{2,2}f_2+w_{2,3}f_2 = (w_{2,2}+w_{2,3})f_2 = f_2$. We also recover $w_{2,2} = x_2/f_2$ and $w_{2,3} = x_3/f_2$.
\end{remark}

%%%%%%%%%%%%%%%%%%%%%%%%%%%%%%%%%
%
\begin{remark}
In the measurement equation (\ref{eq:conservation_law_matrix_form_1}) the unknown weights $w_{k,n}$ are  incorporated into $\vc{x}$. This is  different from  studies in which these weights are known a priori.
Our approach increases the dimension of the unknown vector $\vc{x}$. However, when $\vc{x}$ is suffiiciently sparse, it can be recovered by CODE.  Of course, if some
flows and weights are known, these values can replace the corresponding variables.
\end{remark}

%%%%%%%%%%%%%%%%%%%%%%%%%%%%%%%%%
%

\subsection{Dynamic Traffic Model}
\label{sec:dynamicModel}
In a dynamic  model,  the \ac{OD} flows $f_k$ and link counts $y^i_j$ are time-dependent,  e.g., hourly.  More significantly, 
we need to account for the time delay between the \emph{start time} of a vehicle's trip and the \emph{count time} when its presence on a link is measured.  To illustrate how the measurement equation \eqref{eq:conservation_law_matrix_form_1}  changes, we consider the network  of Fig.~\ref{fig:network_model},
assuming it takes one unit time to traverse each link, so that
\begin{equation}
		\left\{ \begin{array}{ll} 
%		& \hspace{-0.45in} \textrm{dynamic link traffic counts:} \\
%		& \\
			y^1_2(t)  &=  f_1(t) + w_{2,2}f_2(t) + f_6(t-1),\\
                y^1_3(t)  &= w_{2,3}f_2(t),\\
                y^2_3(t) &= w_{2,2}f_2(t-1) + f_3(t) + f_4(t),\\
                y^3_1(t) &= f_3(t-1)+f_5(t)+f_6(t).
			\end{array} \right.
			\label{eq:small_network_dynamic_linkflows}
\end{equation}
In matrix form 
 (\ref{eq:small_network_dynamic_linkflows}) is written as
\begin{equation}
\vc{y}(t) =
\underbrace{
\begin{bmatrix}
1 & 1 & 0 & 0 & 0 & 0 & 0 & 0 & 0 & 1\\
0 & 0 & 1 & 0 & 0 & 0 & 0 & 0 & 0 & 0\\
0 & 0 & 0 & 1 & 1 & 0 & 1 & 0 & 0 & 0\\
0 & 0 & 0 & 0 & 0 & 1 & 0 & 1 & 1 & 0\\
\end{bmatrix}
}_{A}
\underbrace{
\begin{bmatrix}
f_1(t)\\
w_{2,2}f_2(t)\\
w_{2,3}f_2(t)\\
w_{2,2}f_2(t-1)\\
f_3(t)\\
f_3(t-1)\\
f_4(t)\\
f_5(t)\\
f_6(t)\\
f_6(t-1)
\end{bmatrix}
}_{\vc{x}(t)}
\label{eq:small_network_dynamic}
\end{equation}
where $\vc{y}(t) = [y_2^1(t), y_3^1(t),y_3^2(t),y_1^3(t)$.
Evidently, considering time delays increases the dimension of the unknown $\vc{x}(t)$. But if a time-series  is observed there will be more measurements as well.

\section{Compressive Sensing (CS)}
\label{sec:cs}
\ac{CS} techniques \cite{candes2006robust, donoho2006compressed} are used to recover an unknown signal
$\vc{x} \in \real^N$ from  observations $\vc{y} = A\vc{x} \in \real^M$  ($M \ll N$) when $x$ is sparse, i.e. the number of  
 non-zero entries of $\vc{x}$,  $S \ll N$. 
$S := \|\vc{x}\|_0$ denotes the sparsity level of $\vc{x}$. 
Since $M < N$ there are infinitely many candidate solutions to $\vc{y} = A\vc{x}$ for a given $\vc{y}$.
Recovery of $\vc{x}$ is nonetheless possible if the true signal is sparse.  The recovery algorithm seeks a sparse solution among the candidates.

%%%%%%%%%%%%%%%%%%%%%%%%%%%%%%%%%%%%%%%%%%%%%%%%%%%%%%%%%%%%%%%%%%%%%%%%%%%
\subsection{Recovery via $\ell_0$-minimization}

Recovery of a sparse $\vc{x}$ can be formulated via $\ell_0$-minimization
\begin{equation}
\widehat{\vc{x}}_{\ell_0} := \arg \min \|\vc{x}\|_0 \ \ \ \ \ 
\text{subject to} \ \ \ \ \ \vc{y} = A\vc{x}.
\label{eq:l_0_opt1}
\end{equation}
Problem~(\ref{eq:l_0_opt1}) can be interpreted as finding an $S$-term approximation to $\vc{y}$ given $A$~\cite{donoho2006compressed}. 

\subsection{Recovery via $\ell_1$-minimization}

The $\ell_0$-minimization problem~(\ref{eq:l_0_opt1}) is NP-hard. Results of \ac{CS} indicate that it is not always necessary to solve the 
$\ell_0$-minimization problem~(\ref{eq:l_0_opt1}) to recover $\vc{x}$, and a much simpler problem often yields an equivalent solution: we only need to find the ``$\ell_1$-sparsest'' $\vc{x}$ by solving
\begin{equation}
\widehat{\vc{x}}_{\ell_1} := \arg \min \|\vc{x}\|_1 \ \ \ \ \ 
\text{subject to} \ \ \ \ \  \vc{y} = A\vc{x}.
\label{eq:l_1_opt1}
\end{equation}
The $\ell_1$-minimization problem~(\ref{eq:l_1_opt1}), called Basis Pursuit~\cite{chen1998atomic}, is much simpler and  can be solved as a linear program whose computational complexity is polynomial in $N$.  A `noise-aware' version of the $\ell_1$-minimization (\ref{eq:l_1_opt1}) relaxes the equality constraint as
\begin{equation}
\widehat{\vc{x}}_{\ell_1} := \arg \min \|\vc{x}\|_1 \ \ \ \ \ \text{subject to} \ \ \ \ \ \|\vc{y} - A \vc{x}\|_2 \leq \delta,
\label{eq:l_1_opt1_noise}
\end{equation}
where $\delta$ is a parameter that should increase with measurement noise.
%%%%%%%%%%%%%%%%%%%%%%%%%%%%%%%%%%%%%%%%%%%%%%%%%%%%%%%%%%%%%%%%%%%%%%%%%%%
\subsection{$\ell_0/\ell_1$ Equivalence and the Restricted Isometry Property}
\label{subsec:RIP}

Of course $\widehat{\vc{x}}_{\ell_1}$ is not equal to $\widehat{\vc{x}}_{\ell_0}$ without conditions on  $A$. The \ac{RIP} \cite{candes2005decoding,candes2008people} guarantees that the  $\ell_1$-minimizing solution is equivalent to the  $\ell_0$-minimizing solution. 
But \ac{RIP}  is only a \emph{sufficient}  condition and frequently the $\ell_1$-minimization recovers the true sparse solution even without \ac{RIP}.   The gap between  existing recovery guarantees and  actual recovery performance is being narrowed in different applications.  In particular, when sparse dynamical systems are involved this gap is usually large and has been  investigated in system identification~\cite{ohlsson2010segmentation,toth2011csi,sanandaji2012thesis,shah2012linear,sanandaji2012tutorial}, observability and control  of linear systems~\cite{sanandaji2013observability,zhao2012stability}, and identification of interconnected networks~\cite{sanandaji2011cti,pan2012reconstruction}.

\section{A Traffic Network Example}
\label{sec:synthetic}
\begin{figure}[tb]
\centering
\includegraphics[width=0.95\linewidth]{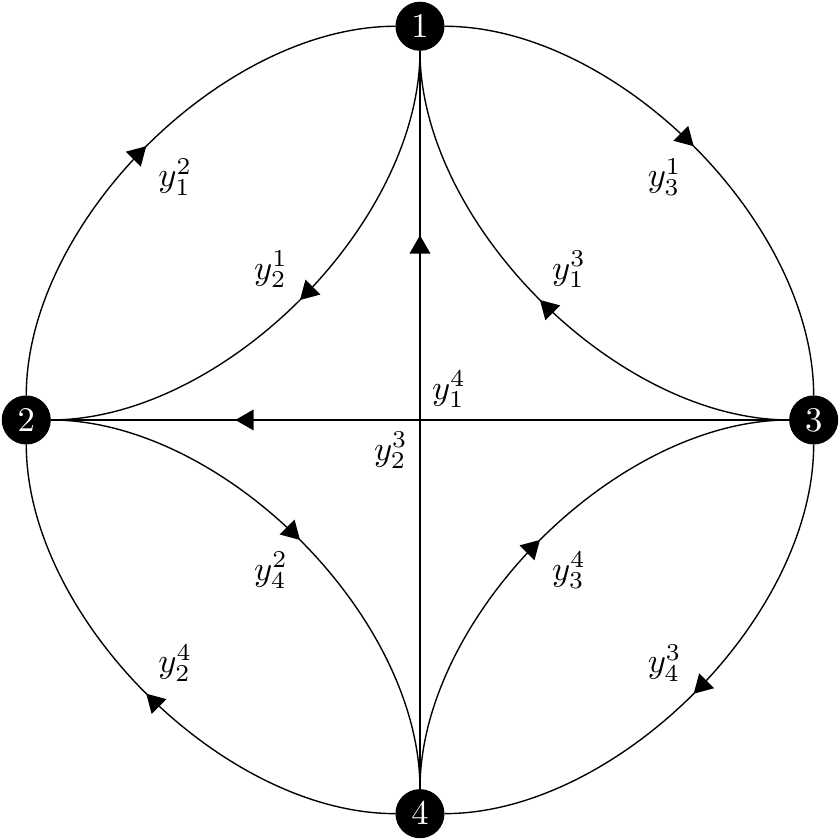}
\caption{Traffic network  with 4 \ac{OD} zones and 10 links.}
\label{fig:transportation_model_medium_1}
\end{figure}

To explain the ideas, consider the static model for the network of  Figure~\ref{fig:transportation_model_medium_1} with $4$ nodes (\ac{OD} zones) and 10 links. Each of the 12 possible \ac{OD} pairs has alternative paths. We assume that only 3 of the \ac{OD} pairs are plausible. Table~\ref{tab:OD-flows} lists these 3 OD pairs and their corresponding 14 paths. When all 10 link counts are available the measurement equation is
\[\vc{y} = \]
\begin{table}
\footnotesize{
\caption{The 3 plausible OD pairs and their corresponding 14 paths for the network of Fig.~\ref{fig:transportation_model_medium_1}.}
\[
\begin{array}{c||c||c|c||c|c}
\hline
\hline
& \color{blue}{\text{OD pair}} \ \ 3-1 &
& \color{blue}{\text{OD pair}} \ \ 3-2 &
& \color{blue}{\text{OD pair}} \ \ 4-2\\
\hline
 \color{blue}{p_1} & \ell^3_1&
 \color{blue}{p_6} & \ell^3_1 \to \ell^1_2 &
 \color{blue}{p_{10}} & \ell^4_1 \to \ell^1_2  \\
  \hline
   \color{blue}{p_2} & \ell^3_2 \to \ell^2_1 &
   \color{blue}{p_7} & \ell^3_2  &
   \color{blue}{p_{11}} & \ell^4_1 \to \ell^1_3 \to \ell^3_2 \\
    \hline
   \color{blue}{p_3} & \ell^3_2 \to \ell^2_4 \to \ell^4_1 &
   \color{blue}{p_8} & \ell^3_4 \to \ell^4_1 \to \ell^1_2&
   \color{blue}{p_{12}} & \ell^4_2  \\
 \hline
   \color{blue}{p_4} & \ell^3_4 \to \ell^4_1 &
   \color{blue}{p_9} & \ell^3_4 \to \ell^4_2 &
   \color{blue}{p_{13}} & \ell^4_3 \to \ell^3_1 \to \ell^1_2  \\
 \hline
   \color{blue}{p_5} & \ell^3_4 \to \ell^4_2 \to \ell^2_1 &
    &  &
   \color{blue}{p_{14}} &  \ell^4_3 \to \ell^3_2  \\
\hline
  \hline
 \end{array}
 \]
 \label{tab:OD-flows}
 }
 \end{table}
\begin{equation}
\hspace{-0.1in}
\underbrace{
\begin{bmatrix}
0  &   0  &   0  &   0  &   0  &   1  &   0  &   1  &   0  &   1  &  0  &   0  &   1  &   0\\
0  &   0  &   0  &   0  &   0  &   0  &   0  &   0  &   0  &   0  &  1  &   0  &   0  &   0\\
0  &   1  &   0  &   0  &   1  &   0  &   0  &   0  &   0  &   0  &  0  &   0  &   0  &   0\\
0  &   0  &   1  &   0  &   0   &  0  &   0  &   0  &   0   &  0  &   0  &   0  &  0  &   0\\
1  &   0  &   0  &   0  &   0   &  1  &   0  &   0  &   0   &  0  &   0  &   0  &   1 &    0\\  
0  &   1  &   1  &   0  &   0  &   0  &   1  &   0  &   0  &   0  &  1  &   0  &   0  &   1\\
0  &   0  &   0  &   1  &   1  &   0  &   0  &   1  &   1  &   0  &  0  &   0  &   0  &   0\\
0  &   0  &   1  &   1  &   0  &   0  &   0  &   1  &   0  &   1  &  1  &   0  &   0  &   0\\
0  &   0  &   0   &  0   &  1  &   0  &   0  &   0  &   1  &   0  &   0  &   1  &   0 &    0\\
0  &   0  &   0  &   0  &   0  &   0   &  0  &   0  &   0  &   0  &   0  &   0  &   1  &   1\\
\end{bmatrix}
}_{A \in \real^{10 \times 14}}
\underbrace{
\begin{bmatrix}
w_{1,1}f_1\\
w_{1,2}f_1\\
w_{1,3}f_1\\
w_{1,4}f_1\\
w_{1,5}f_1\\
w_{2,6}f_2\\
w_{2,7}f_2\\
w_{2,8}f_2\\
w_{2,9}f_2\\
w_{3,10}f_3\\
w_{3,11}f_3\\
w_{3,12}f_3\\
w_{3,13}f_3\\
w_{3,14}f_3\\
\end{bmatrix}
}_{\vc{x} \in \real^{14}}.
\label{eq:matrix_example_1}
\end{equation}
where $\vc{y} = [y_2^1,y_3^1,y_1^2,y_4^2,y_1^3,y_2^3,y_4^3,y_1^4,y_2^4,y_3^4] \in \real^{10}$.

One common way to recover $\vc{x}$ is via  $\ell_2$-minimization:
\begin{equation}
\widehat{\vc{x}}_{\ell_2} := \arg \min \|\vc{x}\|_2 \ \ \ \ \ \text{subject to} \ \ \ \left\{ \begin{array}{cc} 
			\vc{y} = A\vc{x} \\
			x_i \geq 0, \forall i.
			\end{array} \right.
\label{eq:l_2_opt_positive}
\end{equation}
This is one of several `least-squares'  estimation techniques studied in the  literature \cite{abrahamsson1998estimation,bert2009thesis,bera2011estimation}.  We will compare $\widehat{\vc{x}}_{\ell_2}$ and  $\widehat{\vc{x}}_{\ell_1} $:
\begin{equation}
\widehat{\vc{x}}_{\ell_1} := \arg \min \|\vc{x}\|_1 \ \ \ \ \ \text{subject to} \ \ \ \left\{ \begin{array}{cc} 
			\vc{y} = A\vc{x} \\
			x_i \geq 0, \forall i.
			\end{array} \right.
\label{eq:l_1_opt_positive}
\end{equation}		
The constraint $x_i \geq 0, \forall i$ ensures that the path allocations are non-negative.  (Such non-negativity constraints are not usually present in \ac{CS}.)
\subsection{Noiseless Recovery of Sparse Path Allocations}
We consider several scenarios in which  $\ell_1$-minimization (\ref{eq:l_1_opt_positive}) successfully recovers $\vc{x}$ for the network of  Fig.~\ref{fig:transportation_model_medium_1}.
\begin{example}[$\ell_1$-Recovery vs. $\ell_2$-Recovery]
\label{exm:l1-l2}
%Fig.~\ref{fig:transportation_model_medium_1
Suppose we have only $6$ link measurements: $\{y^1_2,y^1_3,y^2_1,y^3_2,y^3_4,y^4_3\}$. The goal is to recover $\vc{x} \in \real^{14}$ based on these measurements. The true path allocation $\vc{x}$ is $4$-sparse. Specifically $p_2$ is used by \ac{OD} pair $3-1$ with $w_{1,2} = 1$, $p_8$  by \ac{OD} pair $3-2$ with $w_{2,8} = 1$, and $p_{11}$ and $p_{14}$ are  taken by \ac{OD} pair $4-2$ with $w_{3,11} = 0.25$ and $w_{3,14} = 0.75$.  Fig.~\ref{fig:transportation_recovery_medium_1} depicts the recovery results:  $\ell_1$-minimization recovers the true $4$-sparse $\vc{x} \in \real^{14}$ from the $6$ link measurements while  $\ell_2$-minimization fails, which motivates \ac{CODE}. As expected  the least squares estimate is not  sparse.
%\hfill $\square$
%
\begin{figure}[tb]
\centering
\includegraphics[width=1\linewidth]{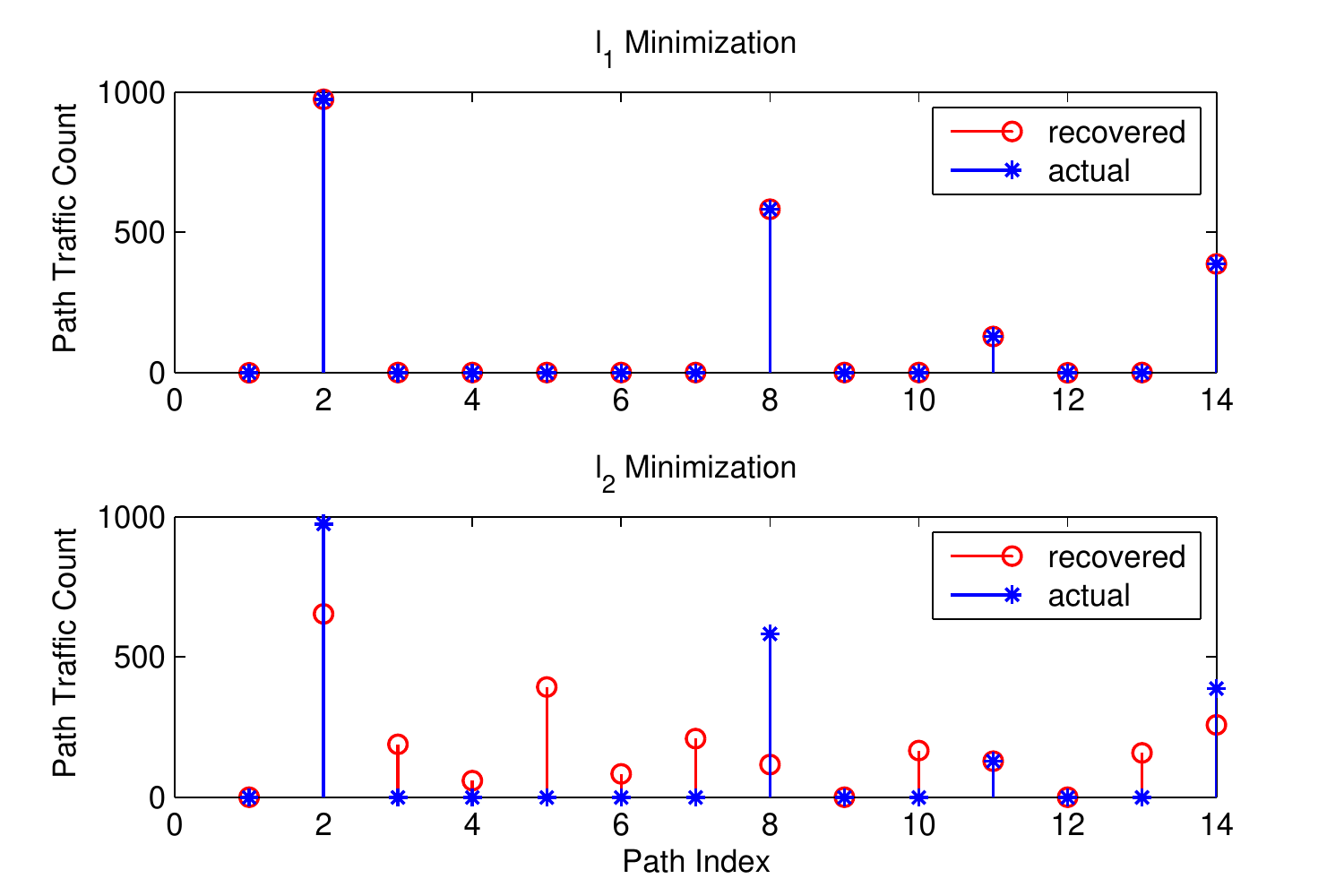}
\caption{(Example \ref{exm:l1-l2}) Illustration of how  $\ell_1$-minimization succeeds and $\ell_2$-minimization fails to recover the true $4$-sparse path allocation from 6 link flow measurements of the traffic network of Fig.~\ref{fig:transportation_model_medium_1}.
% (a) $\ell_1$ recovery. (b) $\ell_2$ recovery.
}
\label{fig:transportation_recovery_medium_1}
\end{figure}
\end{example}

%%%%%%%%%%%%%%%%%%%%%%%%%%%%%%%%%%%%%%%%%%%%%%%%%%%%%%%%%%%%%%%%%%%%%%

\begin{example}
\label{exm:fixedsupport}
We show that  the required number of measurements for exact recovery via $\ell_1$-minimization increases with the sparsity level of $\vc{x}$. Consider two fixed supports for $\vc{x}$: a 3-sparse $\vc{x}$ ($\mathcal{S}_1 = \{5,9,13\}$) and a 4-sparse $\vc{x}$ ($\mathcal{S}_2 = \{2,8,11,14\}$).
For each support, we generate several $\vc{x}$ satisfying (\ref{eq:constraints_on_weights}), and repeat this experiment for different number of link flow measurements. For each fixed number of measurements, we randomly choose a subset of  link flow measurements. For each of 500 trials we solve the $\ell_1$-minimization (\ref{eq:l_1_opt_positive}). 

In order to better characterize the recovered solutions, we consider three different recovery criteria.  The strictest criterion requires $\widehat{\vc{x}}_{\ell_1} = \vc{x}$, so all  path splits and \ac{OD} flows are perfectly estimated. In the second criterion  $\widehat{\vc{x}}_{\ell_1}$ is a successful recovery if all \ac{OD} flows are perfectly recovered while the path splits may not be correctly estimated. The weakest criterion only requires the sum of all  \ac{OD} flows to match the true total flow sum, i.e. the total number of vehicles is correctly estimated, while the OD flows and path split estimates may have errors. 
These criteria may be appropriate depending on the application. 
Fig.~\ref{fig:transportation_medium_recovery_compare_1} summarizes the recovery results. 
Fig.~\ref{fig:medium_net_setAnalysis_Recovery_x_3sparse_1} and \ref{fig:medium_net_setAnalysis_Recovery_x_4sparse_1} depict how the  performance improves as the recovery criterion is relaxed for a $3$-sparse and a $4$-sparse $\vc{x}$, respectively. As expected, more measurements are needed to recover a $4$-sparse $\vc{x}$ than a $3$-sparse $\vc{x}$.  
%\hfill $\square$

\begin{figure*}[tb]
\centering
\subfigure[]{
   \includegraphics[width = 0.85\columnwidth]{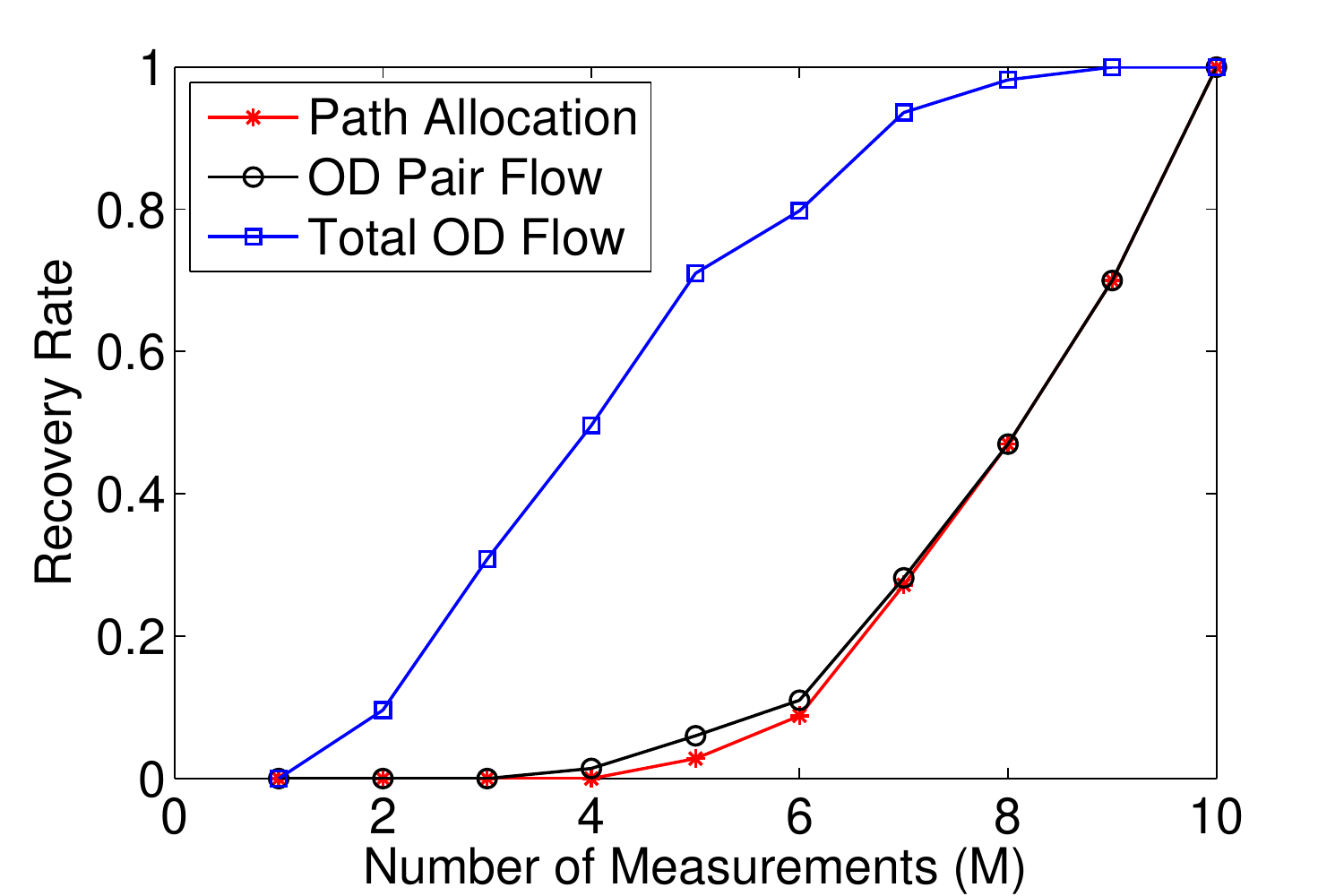}
\label{fig:medium_net_setAnalysis_Recovery_x_3sparse_1}
 }
\subfigure[]{
   \includegraphics[width = 0.85\columnwidth]{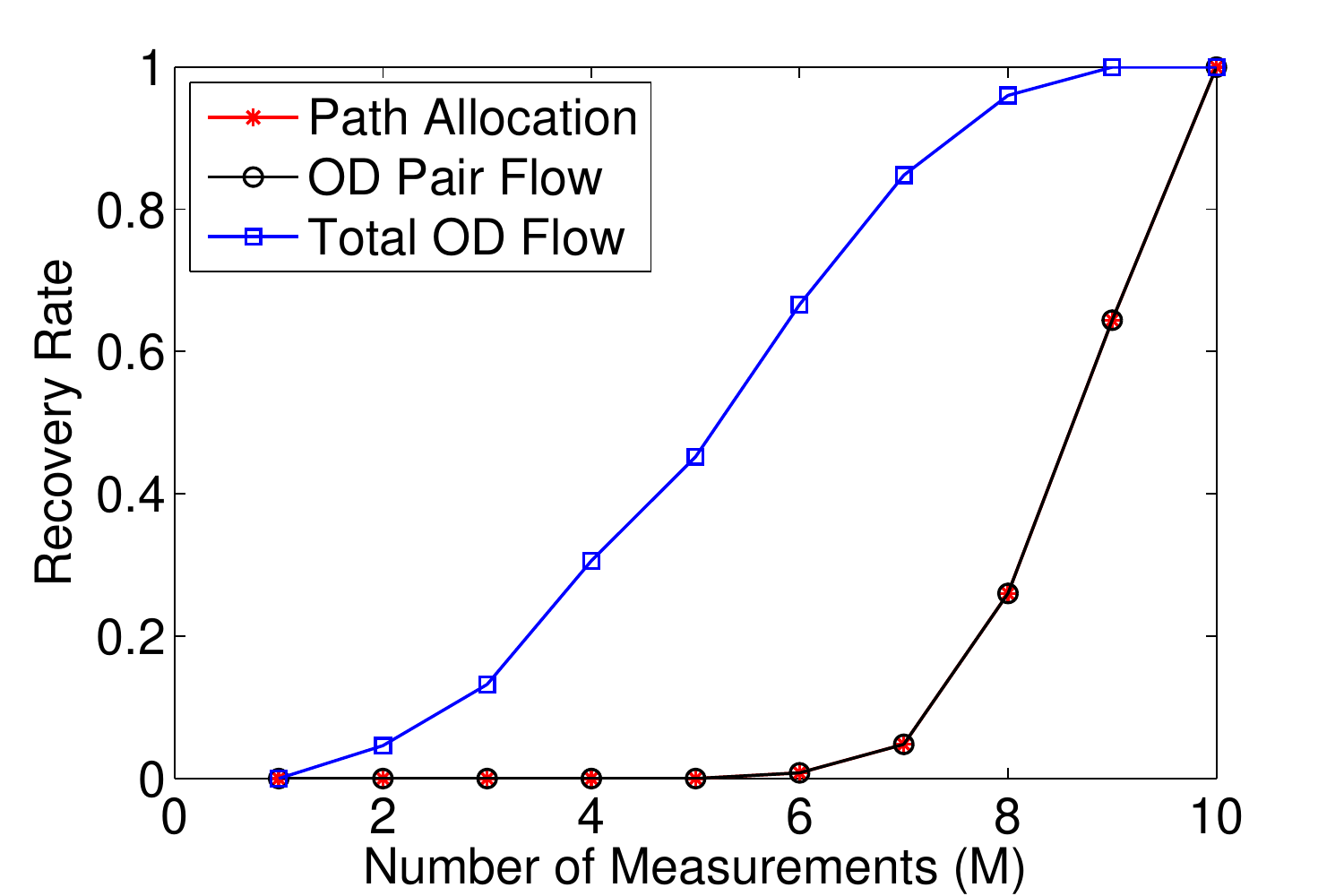}
\label{fig:medium_net_setAnalysis_Recovery_x_4sparse_1}
 }
\caption{(Example~\ref{exm:fixedsupport}) Illustration of how the recovery rate changes with measurements. Two fixed supports (3- and 4-sparse) are considered. In order to better characterize the recovered solutions, three different recovery criteria (path allocation recovery, OD flow recovery, and total OD flow recovery) are considered. (a) 3-sparse path allocation. (b) 4-sparse path allocation.}
\label{fig:transportation_medium_recovery_compare_1}
\end{figure*}

\end{example}

\begin{example}
\label{exm:overSparsity}
We consider all possible $3$-, 4-, and 5-sparse signals and for each, we calculate the recovery rate for different number $M$ of measurements over 500 trials. The results are illustrated in Fig.~\ref{fig:transportation_medium_recovery_x_3_4_5_sparse_y_5_10_compare_1}. At each iteration and for a given sparsity level $S$, we randomly generate a signal $\vc{x}$ (with random OD flow values and  weights while satisfying (\ref{eq:constraints_on_weights}) and on a random support). We then randomly select the link flow measurements for a given $M$. For each pair of $M$ and $S$, we repeat this procedure for 500 iterations and calculate the recovery rate as the fraction of  times that there is an exact recovery (based on the strictest recovery criterion) via $\ell_1$-minimization. 
%\hfill $\square$

\begin{figure}[tb]
\centering
\includegraphics[width=0.95\linewidth]{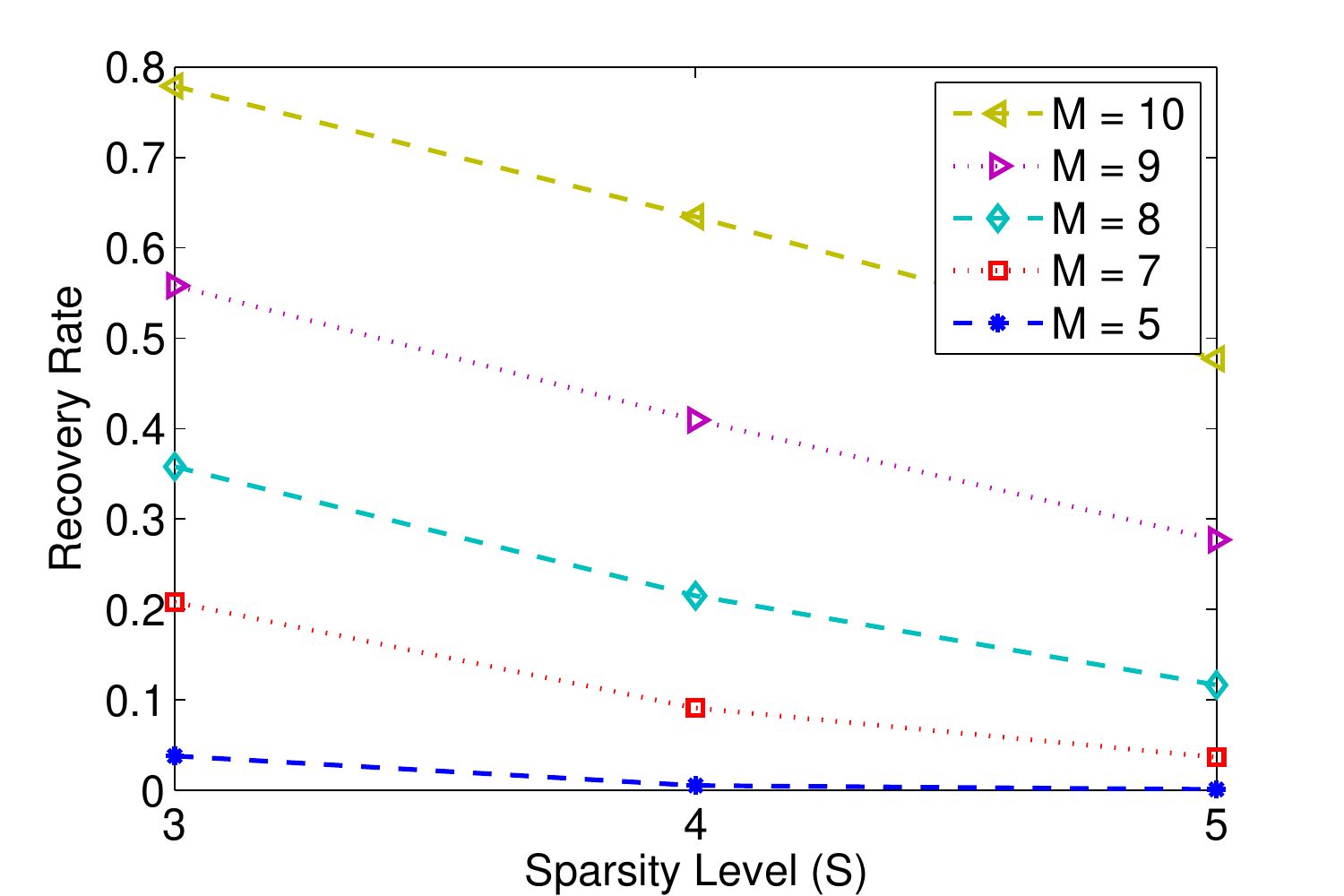}
\caption{(Example \ref{exm:overSparsity}) Illustration of how the recovery rate changes with sparsity and measurements. We perform the following procedure at each iteration. For a given sparsity level and number of measurements, we randomly select the support and generate $\vc{x}$, calculate $\vc{y}$, and then randomly choose a subset of links as our available measurements. For each pair $(S,M)$, we repeat this procedure for 500 iterations and calculate the recovery rate as the fraction of  times there is an exact recovery by solving the $\ell_1$-minimization.}
\label{fig:transportation_medium_recovery_x_3_4_5_sparse_y_5_10_compare_1}
\end{figure}
\end{example}
%%%%%%%%%%%%%%%%%%%%%%%%%%%%%%%%%%%%%%
\subsection{Noisy \ac{CODE} and Compressible Path Allocations}
The ideal case of a sparse signal with noiseless measurements rarely occurs in practice.  One usually deals with \emph{compressible} signals and \textit{noisy} measurements, A signal $\vc{x} \in \real^N$ is said to be compressible when it has more than $S<N$ non-zero entries but can be well approximated by its $S$ largest entries. 

\begin{example}[Noisy $\ell_1$-Recovery]
\label{exm:noisyRecovery}
We illustrate \ac{CODE} for noisy measurements. As in Example \ref{exm:fixedsupport} we consider a 3-sparse and a 4-sparse signal and  $M=10$ link measurements to which are
added Gaussian noises with distribution $\mathcal{N}(0,0.1^2)$ for the 3-sparse signal and distribution $\mathcal{N}(0,0.02^2)$ for the 4-sparse signal. We use a noise-aware version of  (\ref{eq:l_1_opt_positive}):
\begin{equation}
\widehat{\vc{x}}_{\ell_1} := \arg \min \|\vc{x}\|_1 \ \  \text{subject to} \ \ \left\{ \begin{array}{cc} 
			\|\vc{y} - A \vc{x}\|_2 \leq \delta \\
			x_i \geq 0, \forall i.
\end{array} \right.
\label{eq:l_1_opt_positive_noise}
\end{equation}
The parameter $\delta$ depends on the measurement noise.   We compare the estimates of \eqref{eq:l_1_opt_positive_noise} with 
a noise-aware version of the $\ell_2$-minimization (\ref{eq:l_2_opt_positive}):
\begin{equation}
\widehat{\vc{x}}_{\ell_2} := \arg \min \|\vc{x}\|_2 \ \ \text{subject to} \  \ \left\{ \begin{array}{cc} 
			\|\vc{y} - A \vc{x}\|_2 \leq \delta \\
			x_i \geq 0, \forall i.
\end{array} \right.
\label{eq:l_2_opt_positive_noise}
\end{equation}
Fig. \ref{fig:noisyRecovery_x_3sparse_1_y_10_nu_01} and \ref{fig:noisyRecovery_x_4sparse_1_y_10_nu_002} illustrate the noisy recovery results where we plot the empirical \ac{CDF} of recovery error $\|\widehat{\vc{x}}-\vc{x}\|_2/\|\vc{x}\|_2$ over $1000$ iterations, for 3- and 4-sparse signals, respectively. As can be seen,  $\ell_1$-minimization is much less sensitive to noise and has a much smaller recovery error $\|\widehat{\vc{x}}-\vc{x}\|_2/\|\vc{x}\|_2$ in both cases. 
%\hfill $\square$

\begin{figure*}[tb]
\centering
\subfigure[]{
   \includegraphics[width = 0.85\columnwidth]{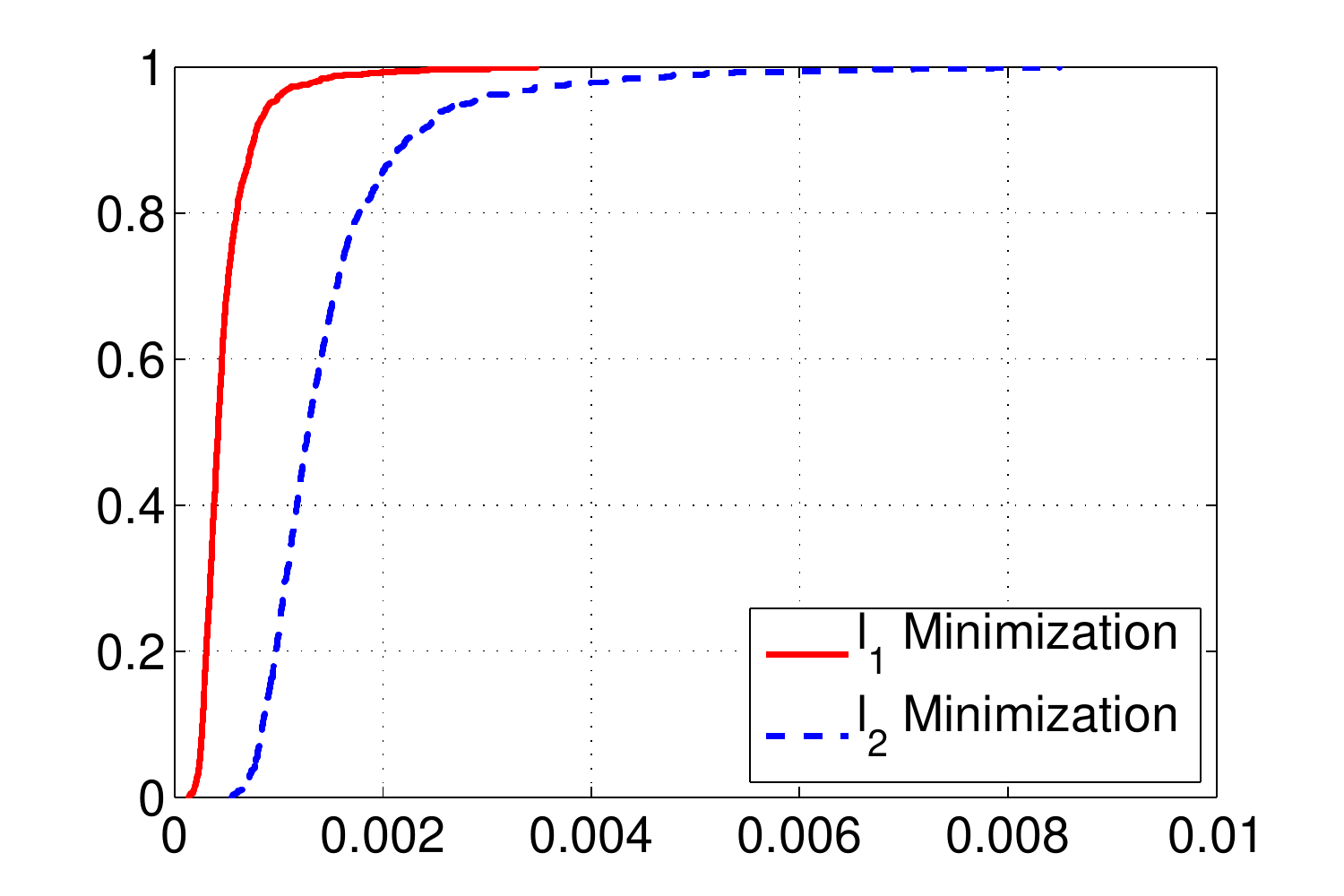}
   \label{fig:noisyRecovery_x_3sparse_1_y_10_nu_01}
 }
\subfigure[]{
   \includegraphics[width = 0.85\columnwidth]{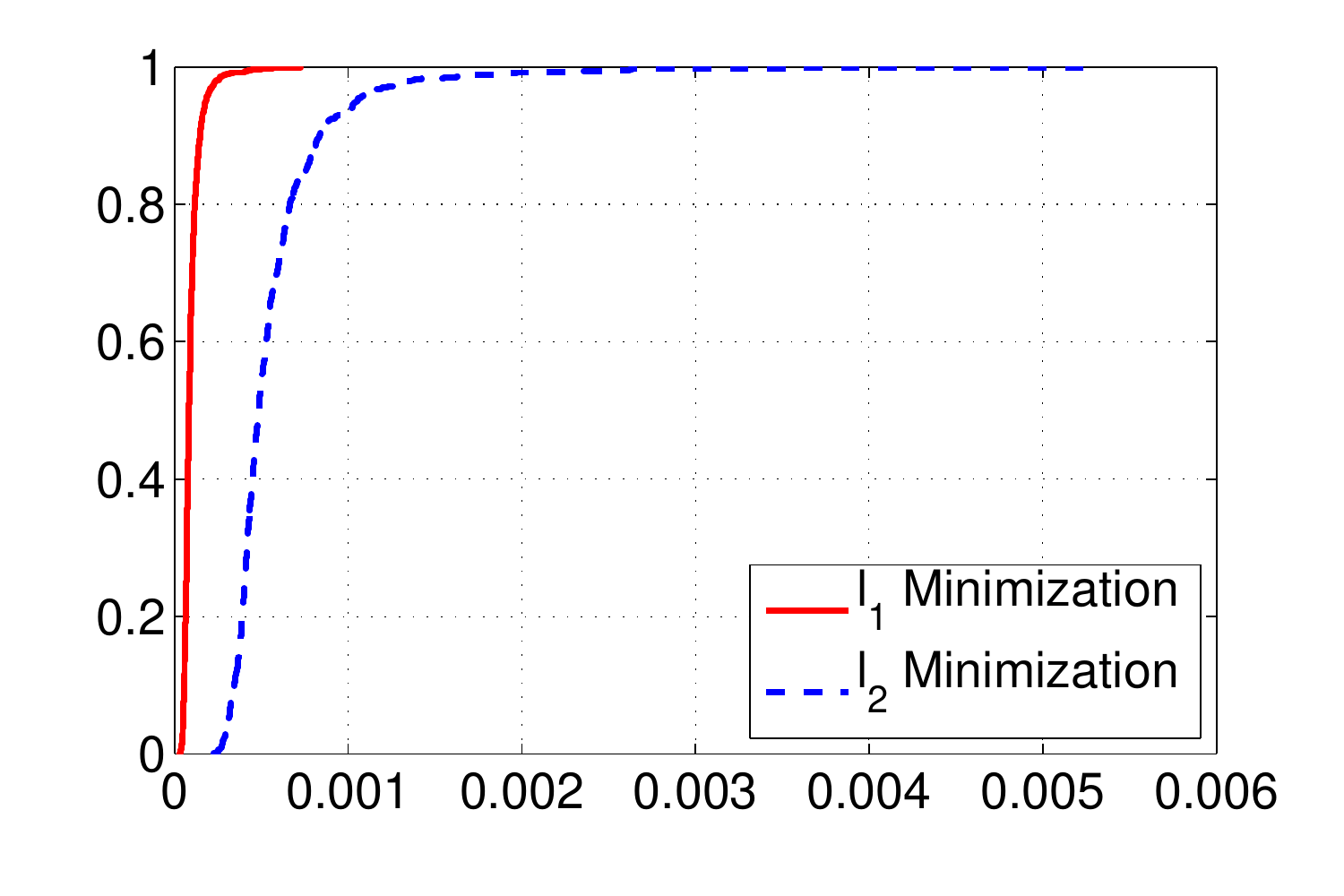}
   \label{fig:noisyRecovery_x_4sparse_1_y_10_nu_002}
 }
 \caption{(Example \ref{exm:noisyRecovery})~\ac{CODE} from noisy measurements. Two fixed supports (a 3-sparse and a 4-sparse) are considered as  in Example \ref{exm:fixedsupport}. All $M=10$ link measurements are available. A Gaussian noise with $\mathcal{N}(0,\nu^2)$ is added to true measurements. We plot the empirical CDF of the error over $1000$ iterations. (a) 3-sparse with $\nu = 0.1$ (b) 4-sparse with $\nu = 0.02$.}
 \label{fig:noisyRecovery_x}
\end{figure*}

\end{example}
%%%%%%%%%%%%%%%%%%%%%%%%%%%%%%%
%%%%%%%%%%%%%%%%%%%%%%%%%%%%%%%
%%%%%%%%%%%%%%%%%%%%%%%%%%%%%%%

%%%%%%%%%%%%%%%%%%%%%%%%%%%%%%%
%%%%%%%%%%%%%%%%%%%%%%%%%%%%%%%

\section{Weighted $\ell_1$-Minimization}
\label{sec:weighted}
Additional knowledge,  for example on AVI or ETC tag data, can improve recovery of the true path allocation vector $\vc{x}$ by using  a \emph{weighted} version of 
the $\ell_1$-minimization problem (\ref{eq:l_1_opt_positive}):

\begin{equation}
\widehat{\vc{x}}_w = \arg \min \|\Lambda \vc{x}\|_1 \ \ \ \ \ \text{subject to} \ \ \ \ \ \left\{ \begin{array}{cc} 
			\vc{y} = A\vc{x} \\
			x_i \geq 0, \forall i,
			\end{array} \right.
\label{eq:l_1_opt1_weighted}
\end{equation}
where $\Lambda \in \real^{N \times N}$ is a given diagonal matrix with positive entries (weights) on the diagonal. Since  $\|\Lambda \vc{x}\|_1 = \sum_{i} \lambda_i|x_i|$ is a weighted sum of the entries of $\vc{x}$,  this is also a linear program.   An entry of $\vc{x}$ assigned  a large weight  gets more penalized in the minimization problem. 
\begin{figure}[tb]
\centering
\includegraphics[width=1\columnwidth]{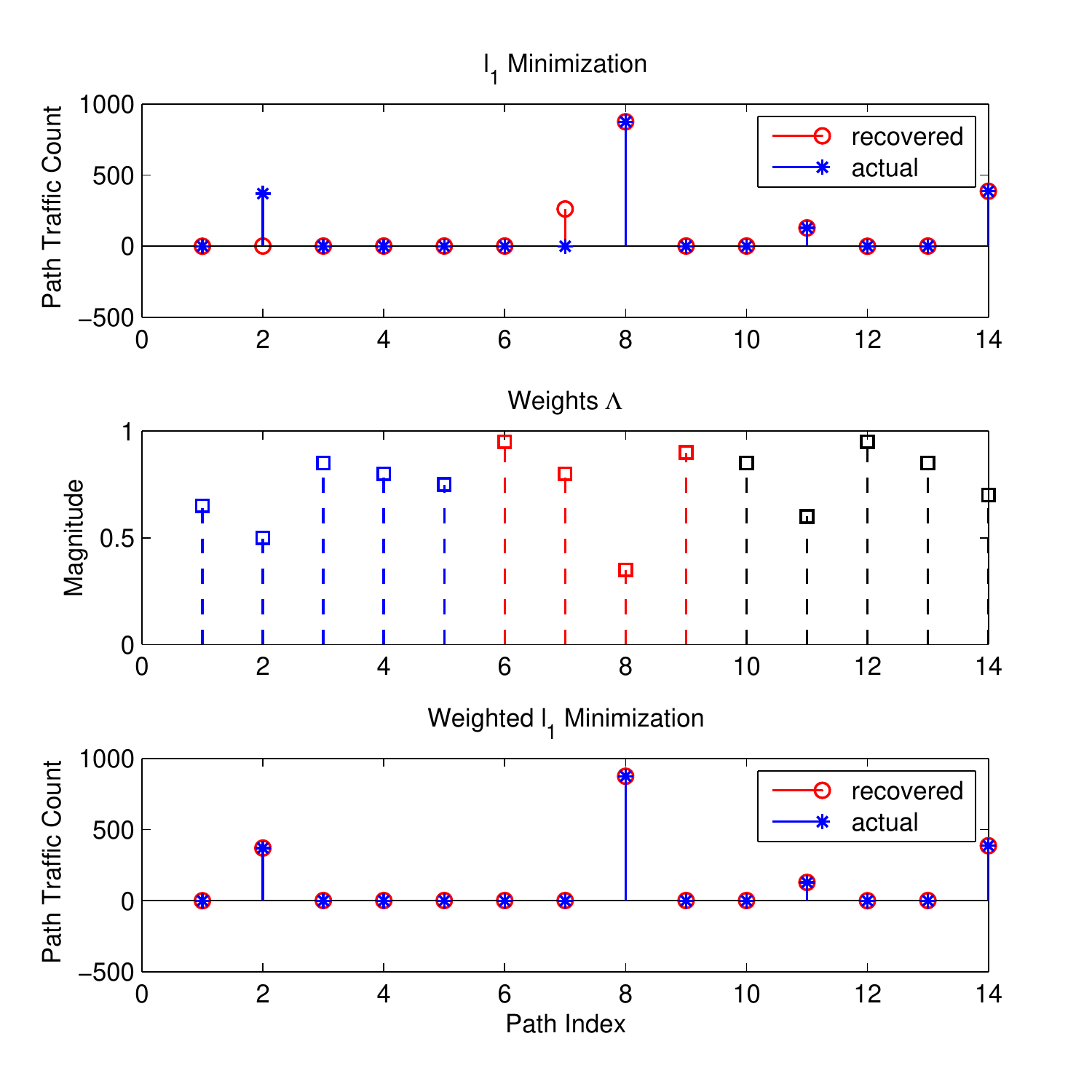}
\caption{Illustration of how weighted $\ell_1$-minimization can help recovery of a $4$-sparse path allocation $\vc{x} \in \real^{14}$ from 6 link flow measurements of the traffic network given in Fig.~\ref{fig:transportation_model_medium_1}.}
\label{fig:medium_net_recovery_4sparse_weighted_1}
\end{figure}
Fig.~\ref{fig:medium_net_recovery_4sparse_weighted_1} shows an example where incorporating  extra information in   \ac{CODE} improves the recovery of a 4-sparse $\vc{x}$ from 6 link flow measurements $\{y^1_2,y^1_3,y^3_2,y^3_4,y^4_2,y^4_3\}$ using weighted $\ell_1$-minimization~(\ref{eq:l_1_opt1_weighted}). 
The weights are chosen based on our prior knowledge of the path allocations. For example, for \ac{OD} pair $3-1$, a smaller weight is assigned to the entry associated with the true path ($p_2$) compared to other alternative paths for this \ac{OD} pair. Similarly, smaller weights are assigned to entries $8$, $11$, and $14$, forcing the weighted $\ell_1$-minimization to penalize these entries less than other paths. 

Ideally, one should assign smaller weights to the non-zero entries associated with the true solution, but we may not know the true support when designing $\Lambda$. In such situations, one can consider an iterated version of $\ell_1$-minimization~\cite{candes2008enhancing}.  At each iteration, the  weight matrix is updated based on the recovered solution at the previous iteration, guided by the extra knowledge. 

%%%%%%%%%%%%%%%%%%%%%%%%%%%%%%
%%%%%%%%%%%%%%%%%%%%%%%%%%%%%%
%%%%%%%%%%%%%%%%%%%%%%%%%%%%%%

\section{Vehicle-Miles Traveled (VMT)}
\label{sec:vmt}

\ac{VMT} is used to estimate  emissions and energy consumption, to allocate resources and assess traffic 
impact~\cite{hoang1980estimating}.  Various methods have been proposed to estimate \ac{VMT} \cite{kumapley1996review}. 
Weighted $\ell_1$-minimization  (\ref{eq:l_1_opt1_weighted}) can also be used to estimate \ac{VMT} using link traffic counts by solving
\begin{equation}
\min_{\vc{x}} \sum_i  v_ix_i\ \ \ \ \ \text{subject to} \ \ \ \ \ \left\{ \begin{array}{cc} 
			\vc{y} = A\vc{x} \\
			x_i \geq 0, \forall i,
			\end{array} \right.
\label{eq:vmt}
\end{equation}
where $v_i$ is the length and $x_i$ is the flow on the $i$th path. 
We also consider a weighted $\ell_1$-maximization problem:
\begin{equation}
\max_{\vc{x}} \sum_i  v_ix_i\ \ \ \ \ \text{subject to} \ \ \ \ \ \left\{ \begin{array}{cc} 
			\vc{y} = A\vc{x} \\
			x_i \geq 0, \forall i.
			\end{array} \right.
\label{eq:vmt_max}
\end{equation}
Observe that the true \ac{VMT} is lower bounded by (\ref{eq:vmt}) and upper bounded by (\ref{eq:vmt_max}). Also, if all $v_i = 1$, we get estimates of the number of vehicles.
\begin{figure}
\begin{center}
 \includegraphics[width = 0.85\columnwidth]{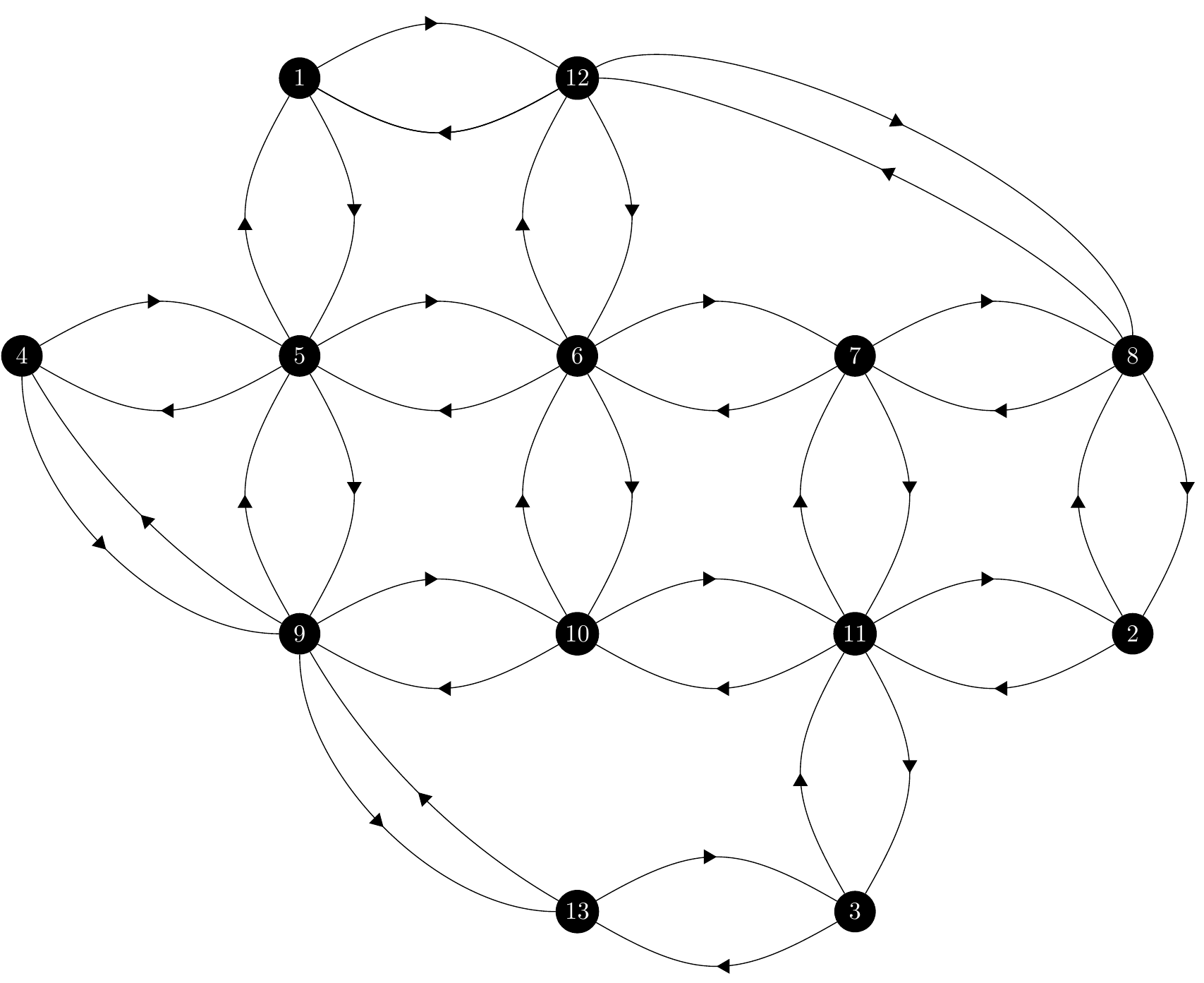}
 \end{center}
 \caption{Nguyen-Dupuis network with 13 nodes and 38 links.}
 \label{fig:nguyen_net}
\end{figure}

%%%%%
\begin{example}[\ac{VMT} Estimation]
We consider the Nguyen-Dupuis network ~\cite{nguyen1984efficient} of Fig. \ref{fig:nguyen_net}. For simplicity,  assume all links  have the same length but different paths have different lengths. There are 8 \ac{OD} plausible pairs: $\{\text{\ac{OD} pair}~1-2,\text{\ac{OD} pair}~1-3,\text{\ac{OD} pair}~2-1,\text{\ac{OD} pair}~2-4,\text{\ac{OD} pair}~3-1,\text{\ac{OD} pair}~3-4,\text{\ac{OD} pair}~4-2,\text{\ac{OD} pair}~4-3\}$, with 50 alternative paths. To save space, we do not list these paths but refer the reader to~\cite{castillo2008observability}. 
We consider a set of $\vc{x} \in \real^{50}$ with 8 non-zero entries, so  there is  one true path for each \ac{OD} pair.
For each measurement we solve~(\ref{eq:vmt}) and (\ref{eq:vmt_max}) and compute the recovery rate for different number of link measurements. We repeat this for 500 trials and consider recovery when 
$\|\widehat{\vc{x}}-\vc{x}\|_2 \leq 0.001$, where $\widehat{\vc{x}}$ is the solution of (\ref{eq:vmt}) or (\ref{eq:vmt_max}).
Fig.~\ref{fig:NguyenNetwork_recovery_1} shows the recovery results. Recovery improves with the number of measurements.    

It is revealing to look at  cases when exact recovery fails. Since a solution to~(\ref{eq:vmt}) is a lower bound to the \ac{VMT}, and a solution to ~(\ref{eq:vmt_max})  is an upper
bound, the ratios 
$\vc{v}^T\widehat{\vc{x}}_{\text{VMT Min}}/\vc{v}^T\vc{x} <1 < \vc{v}^T\widehat{\vc{x}}_{\text{VMT Max}}/\vc{v}^T\vc{x}$  measure the accuracy of the estimates when recovery fails. ($\vc{v}^T$ is the transpose of the vector $\vc{v} = [v_1, v_2, \dots, v_N]^T$.)
Fig.\ref{fig:NguyenNetwork_vmtFailure_1} displays the mean values of the ratios when  exact recovery fails as a function of the number of link measurements.  
 With 22 measurements both (\ref{eq:vmt}) or (\ref{eq:vmt_max}) yield  $80\%$  recovery.  But even in the $20\%$ of the trials that exact recovery fails, the  solutions are within 5 percent of the true value. 

\label{exm:complexityAnalysis}
%\hfill $\square$

\begin{figure*}[tb]
\centering
\subfigure[]{
   \includegraphics[width = 0.85\columnwidth]{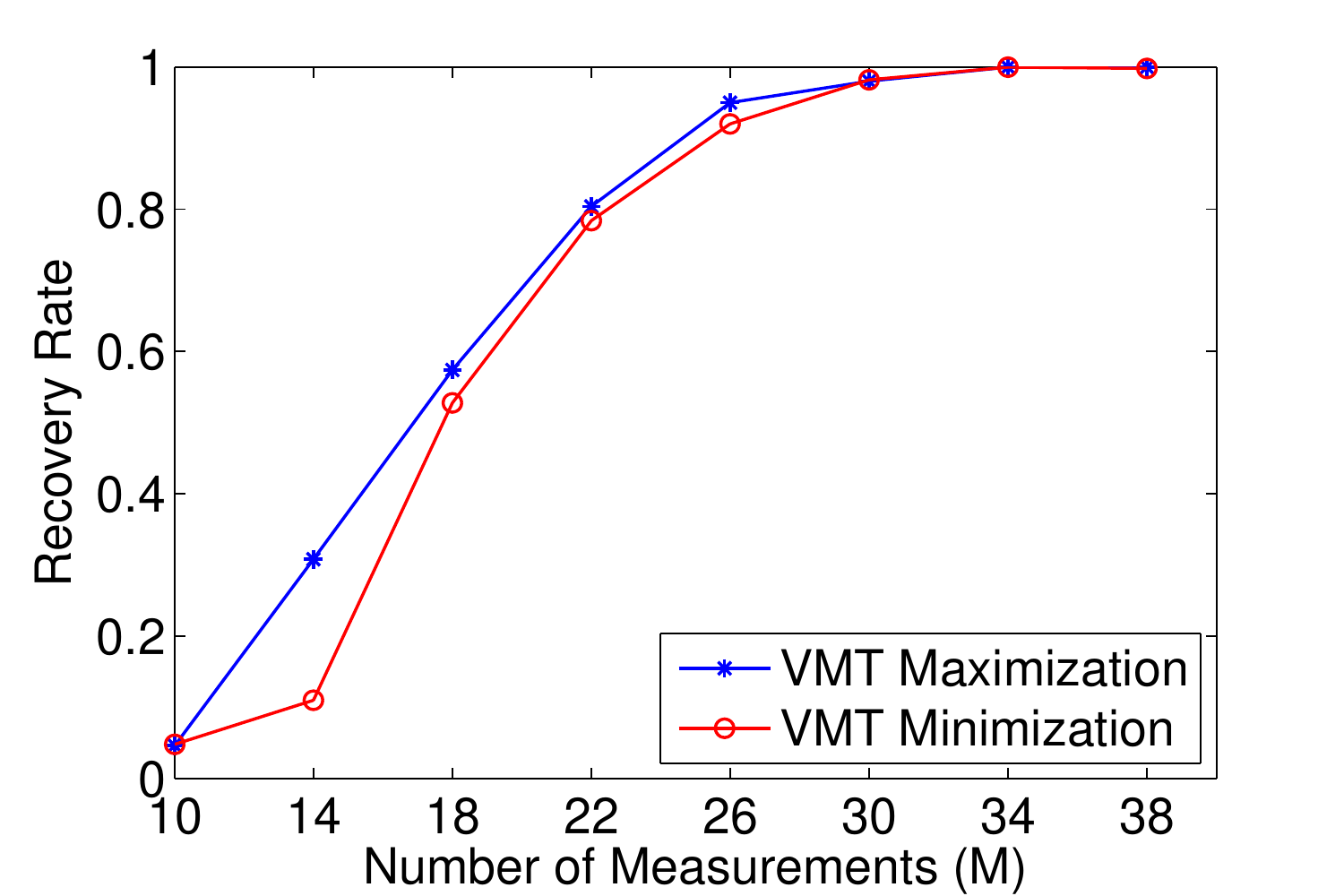}
   \label{fig:NguyenNetwork_recovery_1}
 }
\subfigure[]{
\centering
   \includegraphics[width = 0.85\columnwidth]{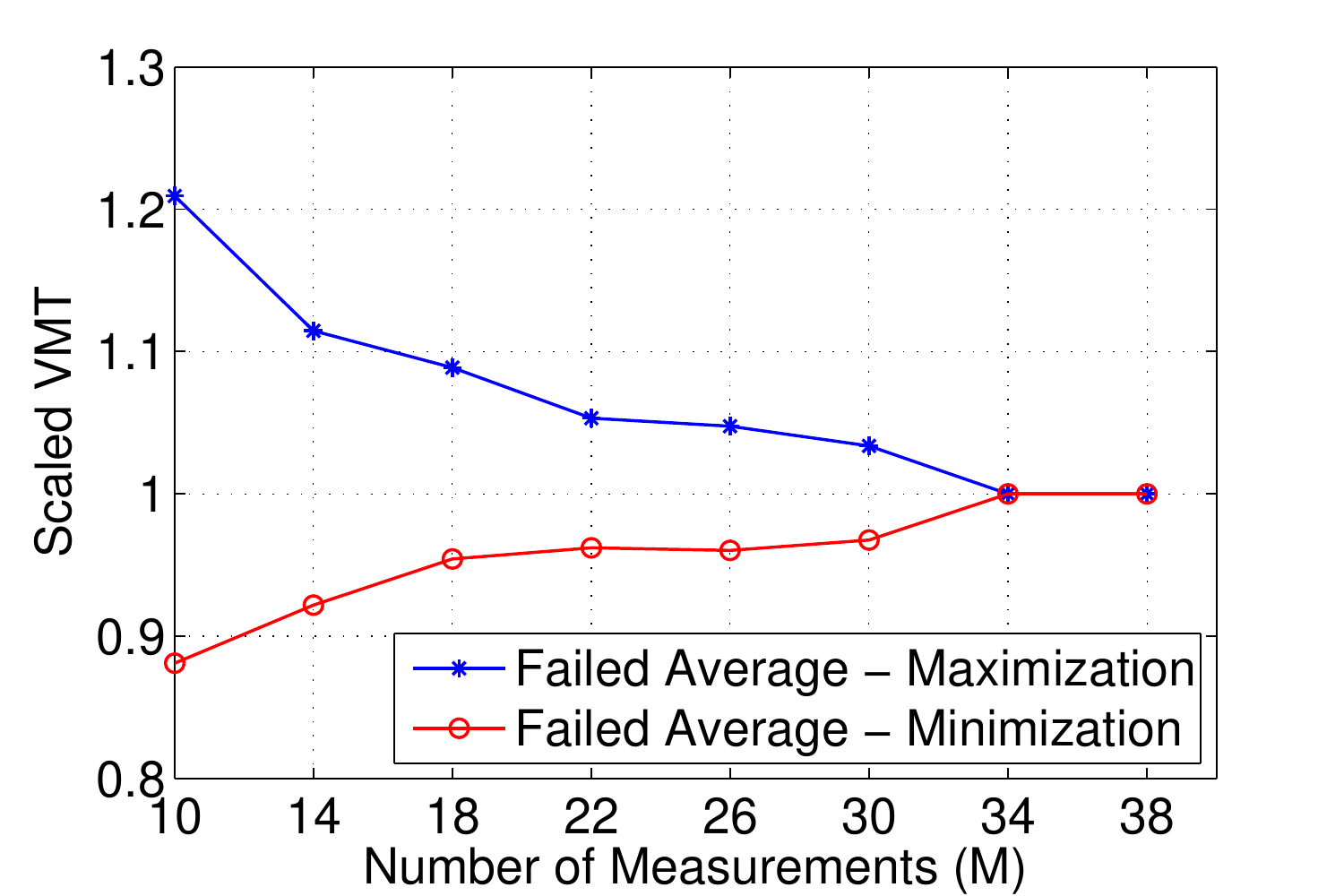}   \label{fig:NguyenNetwork_vmtFailure_1}
 }
 \caption{(Example~\ref{exm:complexityAnalysis}) \ac{VMT} estimation of Nguyen-Dupuis network. (a) Recovery results. For each measurement, we solve~(\ref{eq:vmt}) and (\ref{eq:vmt_max}) and compute the recovery rate for different number of link measurements. We repeat this for 500 trials and consider recovery when 
$\|\sum v_i (\widehat{{x}}_{i}-{x}_i)\|_2 \leq 0.001$ where $\widehat{\vc{x}}$ is the recovered solution solving (\ref{eq:vmt}) or (\ref{eq:vmt_max}).
(b) Mean value of $\vc{v}^T\widehat{\vc{x}}_{\text{VMT Min}}/\vc{v}^T\vc{x}$ and $\vc{v}^T\widehat{\vc{x}}_{\text{VMT Max}}/\vc{v}^T\vc{x}$ when (\ref{eq:vmt}) and (\ref{eq:vmt_max}) fail to exactly recover, respectively.}
\end{figure*}

\end{example}

%%%%%%%%%%%%%%%%%%%%%%%%%%%%%%%
%%%%%%%%%%%%%%%%%%%%%%%%%%%%%%%
%%%%%%%%%%%%%%%%%%%%%%%%%%%%%%%

\section{Case Study in East Providence}
\label{sec:case}
\begin{figure}[tb] 
\centering
\subfigure[]{
   \includegraphics[width = 0.9\columnwidth]{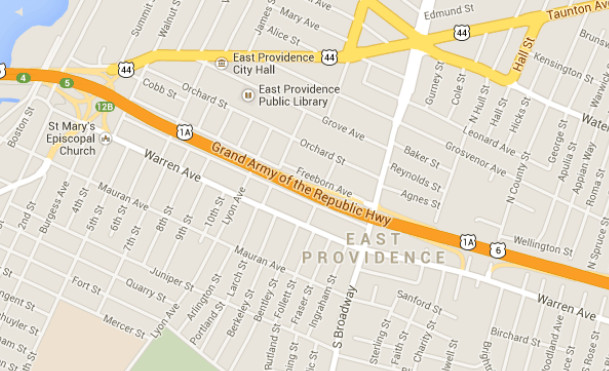}
   \label{fig:map_warren_1}
   }
\subfigure[]{
   \includegraphics[width = 0.9\columnwidth]{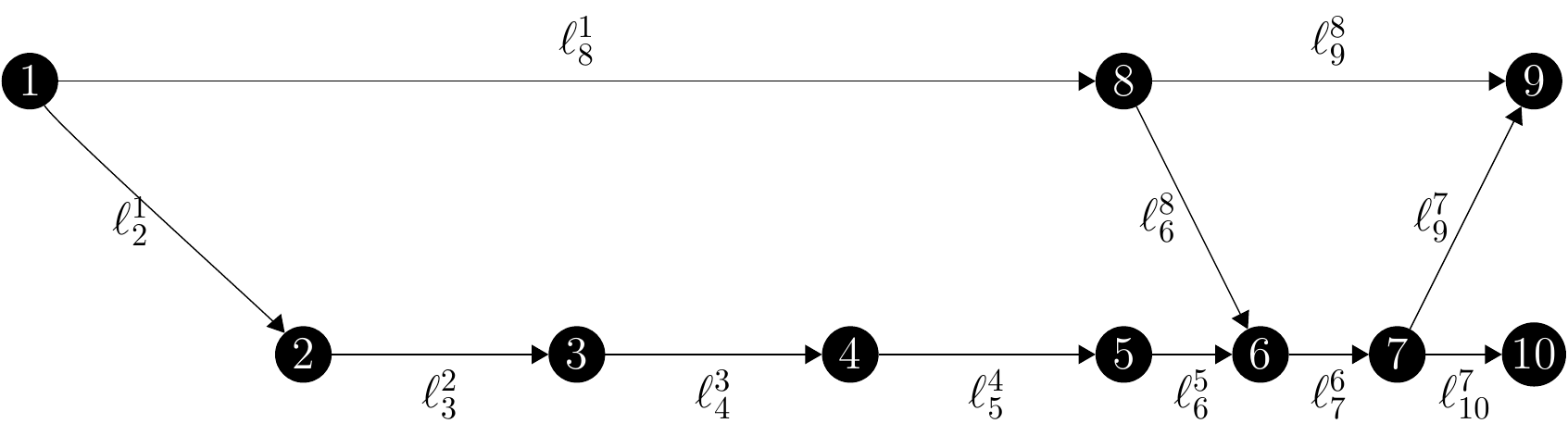}
   \label{fig:sch_warren_1}
   }
\caption{Case study of \ac{CODE} on real data. (a) Map of the area under study in East Providence (Grand Army of the Republic Hwy and Warren Ave). (b) Schematic of the area under study. There are a total of 33 paths associated with \ac{OD} pairs while 10 link flow measurements are available.}
\label{fig:warren_1}
\end{figure}

\begin{figure}[tb]
\centering
\includegraphics[width=0.8\columnwidth]{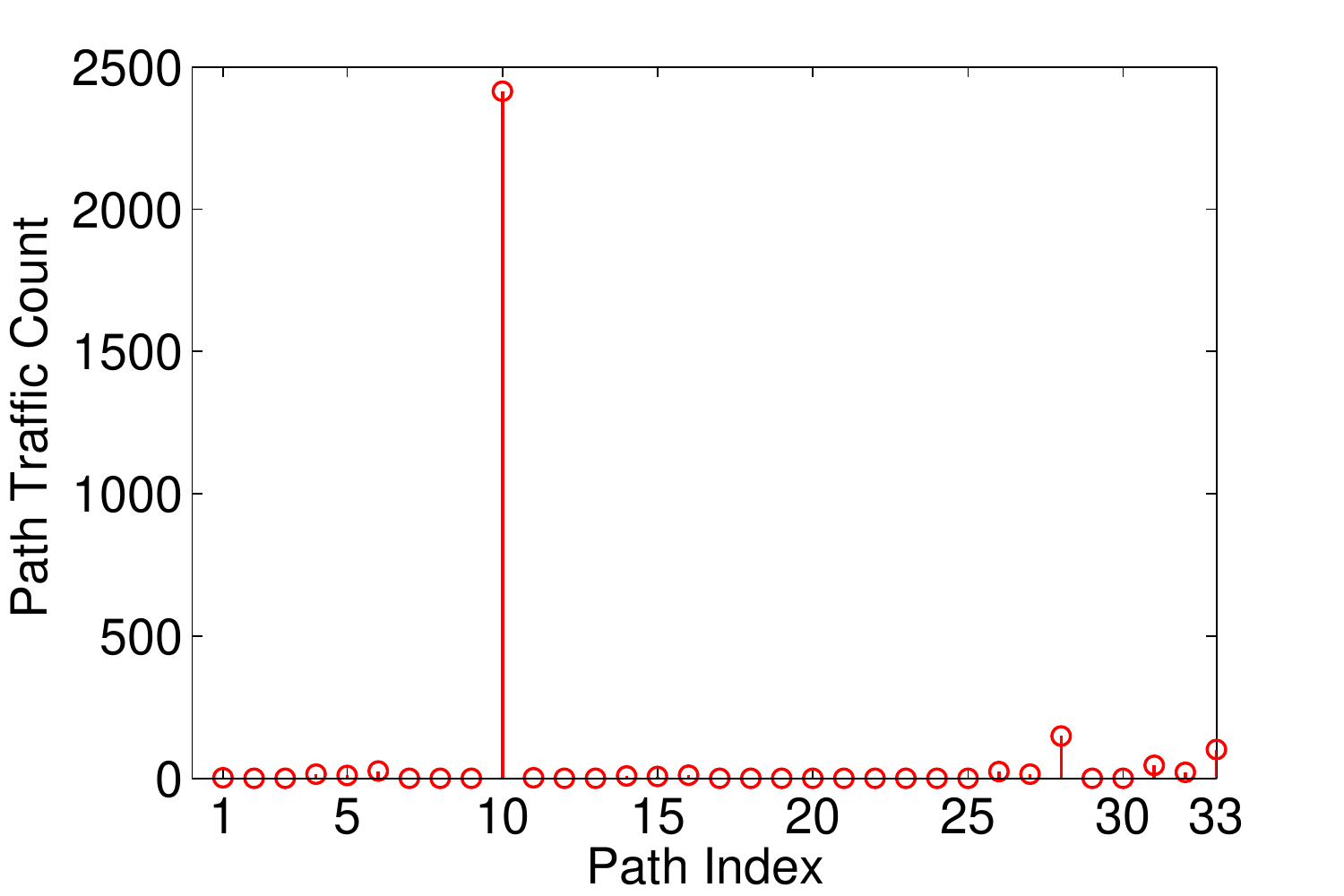}
\caption{The recovered path allocation using \ac{CODE} associated with the considered area in East Providence as shown in Fig.~\ref{fig:map_warren_1} and Fig.~\ref{fig:sch_warren_1}. As can be seen, the solution has only a few non-zero entries. The most significant path is the path on the Grand Army of the Republic Highway ($ \ell^1_8 \to \ell^8_9$).}
\label{fig:warren_sim_1_hour_6_1}
\end{figure}

We apply \ac{CODE} to traffic data recorded from an area in East Providence. Figure~\ref{fig:warren_1} depicts the map and  the network of the  area. The corresponding  $A$ matrix has $10$ rows (number of measurements) and $33$ columns (number of paths).  To save space we do not display $A$ and focus on  the  recovery results summarized in 
Fig.~\ref{fig:warren_sim_1_hour_6_1}.  The solution has only a few non-zero entries. The most significant path ($p_{10}$) is along the Grand Army of the Republic Highway ($ \ell^1_8 \to \ell^8_9$).
This result further confirms that there usually exists a sparse path allocation which can be recovered using \ac{CODE}.

%%%%%%%%%%%%%%%%%%%%%%%%%%%%%%
%%%%%%%%%%%%%%%%%%%%%%%%%%%%%%
% 

%%%%%%%%%%%%%%%%%%%%%%%%%%%%%%%%%%%%%%

\section{Conclusions and Future Work}
We proposed \ac{CODE}, an algorithm to estimate \ac{OD} flows and their path allocations. Three variants of \ac{CODE} (noiseless, noisy, and weighted) were considered, all involving $\ell_1$-minimization. Examples suggest that when the true path allocation is suitably sparse, \ac{CODE}  recovers the unknown variables exactly, even from a highly underdetermined set of linear equations.

Future directions for work include examining large networks to understand better the relation between sparsity, accuracy and computational effort. Also worth investigation is the incorporation of additional \ac{OD} information obtained via Bluetooth or cell phone records. Another interesting question is to use \ac{CODE} to determine path allocations that form a user equilibrium. Yet another direction concerns the selection of additional link counts that improve recovery.

%%%%%%%%%%%%%%%%%%%%%%%%%%%%%%%%%%%%%%

\bibliographystyle{IEEEtran}
\bibliography{reference-traffic.bib}

\begin{IEEEbiography}[{\includegraphics[width=1in,height=1.25in,clip,keepaspectratio]{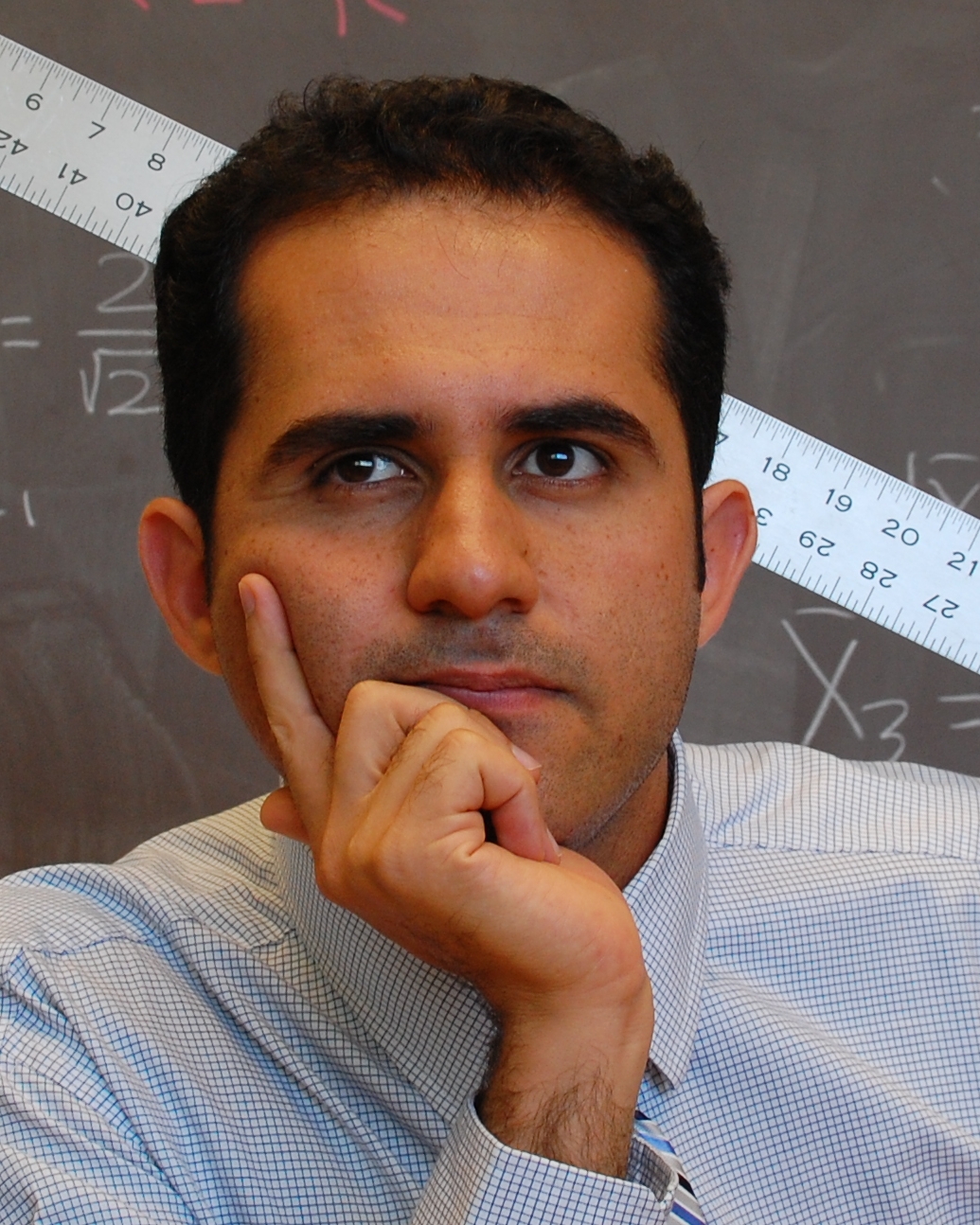}}]{Borhan M. Sanandaji}
is a postdoctoral scholar at the University or California, Berkeley in the Electrical Engineering and Computer Sciences department. He received his Ph.D.
degree (2012) in electrical engineering from the Colorado School of Mines and his B.Sc. degree
(2004) in electrical engineering from the Amirkabir University of Technology (Tehran, Iran). His current research interests include compressive sensing, low-dimensional modeling, and big physical data analytics with applications in energy systems, control, and transportation.
\end{IEEEbiography}

\begin{IEEEbiography}[{\includegraphics[width=1in,height=1.25in,clip,keepaspectratio]{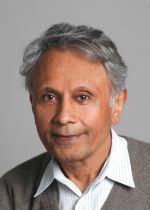}}]{Pravin P. Varaiya}
is Professor of Graduate School in the Dept. of Electrical Engineering and Computer Sciences at the University of California, Berkeley.  His current research
concerns transportation networks and electric
power systems.  He received  the Field
Medal and Bode Prize of the IEEE Control Systems
Society, the Richard E. Bellman Control Heritage
Award, and the Outstanding Research Award of the
IEEE Intelligent Transportation Systems Society. He is a Fellow of IEEE, a member of the National Academy of Engineering, and a Fellow of the American Academy of Arts.
\end{IEEEbiography}

\end{document}